\documentclass[PRL,reprint,amsmath,amssymb,superscriptaddress]{revtex4-1}

\usepackage{graphicx}
\usepackage{dcolumn}
\usepackage{bm}
\usepackage{hyperref}
\usepackage{breakurl}
\usepackage{multirow}
\usepackage{enumitem}

\begin{document}\title{Helicity protected ultrahigh mobility Weyl fermions in NbP}

\author{Zhen Wang}
      \affiliation{Department of Physics, Zhejiang University, Hangzhou 310027, P. R. China}
       \affiliation{State Key Lab of Silicon Materials, Zhejiang University, Hangzhou 310027, P. R. China}
\author{Yi Zheng}
      \email{phyzhengyi@zju.edu.cn}
      \affiliation{Department of Physics, Zhejiang University, Hangzhou 310027, P. R. China}
      \affiliation{Zhejiang California International NanoSystems Institute, Zhejiang University, Hangzhou 310058, P. R. China}
            \affiliation{Collaborative Innovation Centre of Advanced Microstructures, Nanjing 210093, P. R. China}
\author{Zhixuan Shen}
      \affiliation{Department of Physics, Zhejiang University, Hangzhou 310027, P. R. China}
\author{Yunhao Lu}
      \affiliation{State Key Lab of Silicon Materials, Zhejiang University, Hangzhou 310027, P. R. China}
\author{Hanyan Fang}
      \affiliation{Department of Physics, Zhejiang University, Hangzhou 310027, P. R. China}
\author{Feng Sheng}
      \affiliation{Department of Physics, Zhejiang University, Hangzhou 310027, P. R. China}
\author{Yi Zhou}
      \affiliation{Department of Physics, Zhejiang University, Hangzhou 310027, P. R. China}
      \affiliation{Collaborative Innovation Centre of Advanced Microstructures, Nanjing 210093, P. R. China}
\author{Xiaojun Yang}
      \affiliation{Department of Physics, Zhejiang University, Hangzhou 310027, P. R. China}
\author{Yupeng Li}
      \affiliation{Department of Physics, Zhejiang University, Hangzhou 310027, P. R. China}
\author{Chunmu Feng}
      \affiliation{Department of Physics, Zhejiang University, Hangzhou 310027, P. R. China}
\author{Zhu-An Xu}
   \email{zhuan@zju.edu.cn}
      \affiliation{Department of Physics, Zhejiang University, Hangzhou 310027, P. R. China}
      \affiliation{State Key Lab of Silicon Materials, Zhejiang University, Hangzhou 310027, P. R. China}
      \affiliation{Zhejiang California International NanoSystems Institute, Zhejiang University, Hangzhou 310058, P. R. China}
      \affiliation{Collaborative Innovation Centre of Advanced Microstructures, Nanjing 210093, P. R. China}

\date{\today}

\begin{abstract}

Non-centrosymmetric transition metal monopnictides, including TaAs, TaP, NbAs, and NbP, are emergent topological Weyl semimetals (WSMs) hosting exotic relativistic Weyl fermions. In this letter, we elucidate the physical origin of the unprecedented charge carrier mobility of NbP, which can reach $1\times10^{7}$ cm $^{2}$V$^{-1}$s$^{-1}$ at 1.5 K. Angle- and temperature-dependent quantum oscillations, supported by density function theory calculations, reveal that NbP has the coexistence of p- and n-type WSM pockets in the $k_{z}$=1.16$\pi$/c plane (W1-WSM) and in the $k_{z}$=0 plane near the high symmetry points $\Sigma$ (W2-WSM), respectively. Uniquely, each W2-WSM pocket forms a large dumbbell-shaped Fermi surface (FS) enclosing two neighboring Weyl nodes with the opposite chirality. The magneto-transport in NbP is dominated by these highly anisotropic W2-WSM pockets, in which Weyl fermions are well protected from defect backscattering by real spin conservation associated to the chiral nodes. However, with a minimal doping of $\sim$1\% Cr, the mobility of NbP is degraded by more than two order of magnitude, due to the invalid of helicity protection to magnetic impurities. Helicity protected Weyl fermion transport is also manifested in chiral anomaly induced negative magnetoresistance, controlled by the W1-WSM states. In the quantum regime below 10 K, the intervalley scattering time by impurities becomes a large constant, producing the sharp and nearly identical conductivity enhancement at low magnetic field.
\end{abstract}

\maketitle
Topological Weyl semimetals (WSMs) are regarded as the next wonderland in condensed matter physics \cite{WSMWanXG_PRB,WSMFangZ_PRLHgCrSe,WSMDaiX_PRX,NCHasan_WSMTheory} for exploring fascinating quantum phenomena  \cite{ChiralAnomaly_PLB,VishwaMRanomaly_PRX14,ThermalAnomaly_PRB14,HelicalFermiArcs_PRB2,NCommVishwa_FermiArc,MagnetoOptical_PRB}. Unlike Dirac semimetals (DSMs) \cite{DSMCd3As_FangZ_PRB,DSMCdAs_NPOng_NatMat15}, band crossing points in WSMs, \textit{i.e.} Weyl nodes, always appear in pair with opposite chirality, due to the lifting of spin degeneracy by breaking either time reversal symmetry \cite{WSMWanXG_PRB} or inversion symmetry \cite{WSMDaiX_PRX,NCHasan_WSMTheory}. Fermi surfaces (FSs) enclosing the chiral Weyl nodes are characterized by helicity, \textit{i.e.} the spin orientation is either parallel or antiparallel to the momentum. Such helical Weyl fermions are expected to be remarkably robust against non-magnetic disorders, and may lead to novel device concepts for spintronics and quantum computing.

The recent proposed non-centrosymmetric TaAs, TaP, NbAs and NbP, have stimulated immense interests, due to the binary, non-magnetic crystal structure. The existence of Weyl nodes has soon been discovered in TaAs by angle-resolved photoemission spectroscopy (ARPES) \cite{TaAsDingH_ARPESWSM,TaAsDingH_ARPESNode}, and by quantum transport measurements of NMR and a non-trivial Berry's phase ($\Phi_{B}$) of $\pi$ \cite{TaAs_arXiv_Jia,TaAs_arXiv_NMR}. Transport studies of NbAs \cite{NbAs_jpcm} and NbP \cite{NbP_arXiv} also show ultrahigh mobility and non-saturating MR, but no convincing evidence on the existence of Weyl fermions in these two compounds. However, ARPES resolves tadpole-shaped Fermi arcs on the (001) surface of both NbAs \cite{NbAsHasan_ARPES} and NbP \cite{NbP_CPL_FengDL}. It also shows pronounced changes in the electronic structures of NbAs and NbP compared to TaAs \cite{NbAsHasan_ARPES}, mainly due to weaker spin-orbital-coupling (SOC) in the former two and shifting in the band crossing energy relative to the Fermi energy ($E_{F}$). In TaAs, the dominant WSM electron pockets are enclosing the eight pairs of Weyl node 1 (W1) in the $k_{z}$=1.18 $\pi/c$ plane, while WSM pockets surrounding Weyl node 2 (W2) in the $k_{z}$=0 plane near the four $\Sigma$ points are negligible \cite{TaAs_arXiv_Jia,TaAs_arXiv_NMR}. For NbAs, W2 becomes 36 meV below the nearly neutral W1 \cite{NbAsHasan_ARPES}. It is thus expected that charge transport in NbAs and NbP would notably differ from TaAs, because the W2-Weyl cones are highly anisotropic in k-space compared to the relatively isotropic W1-Weyl cones \cite{WSMDaiX_PRX,TaAsDingH_ARPESNode,NbAsHasan_ARPES}. Intriguingly, the carrier mobility of NbP ($5\times10^{6}$ cm$^{2}$V$^{-1}$s$^{-1}$) \cite{NbP_arXiv} is one order of magnitude higher than TaAs and NbAs \cite{TaAs_arXiv_Jia,TaAs_arXiv_NMR,NbAs_jpcm}. Such striking mobility could either be due to significantly lower concentration of lattice disorders in NbP compared to TaAs and NbAs, or indicate a protection mechanism that effectively suppresses the backscattering of charge carriers \cite{DSMCdAs_NPOng_NatMat15,AndoBackscattering_JPSJ98,GeimGr_NatMat07,BerryPhaseRashba_Science,HasanTI_RevModPhy10}.

In this Letter, we unambiguously prove the existence of chiral WSM states in NbP using angle-dependent quantum oscillations of magnetoresistance, Nernst and Seebeck, compared with the density functional theory calculations. We show that the unprecedented mobility of $1\times10^{7}$ cm$^{2}$V$^{-1}$s$^{-1}$ in NbP is indeed rooted in helical Weyl fermions, associated to four unusually large WSM electron pockets near the $\Sigma$ points in the $k_{z}$=0 plane. Each of such large WSM pockets, which are negligibly small in TaAs, is highly anisotropic in k-space and encloses one pair of the W2 nodes. Robust chiral anomaly induced NMR, another quantum signature of WSMs, has also been demonstrated. However, it is originating in the coexistent p-type WSM pockets in the $k_{z}$=1.16$\pi$/c plane. In the quantum degerate regime, helicity protection of Weyl fermions leads to sharp and nearly identical conductivity enhancement when the magnetic field and electric field are applied in parallel.

\begin{figure}[!thb]
\begin{center}
\includegraphics[width=3.5in]{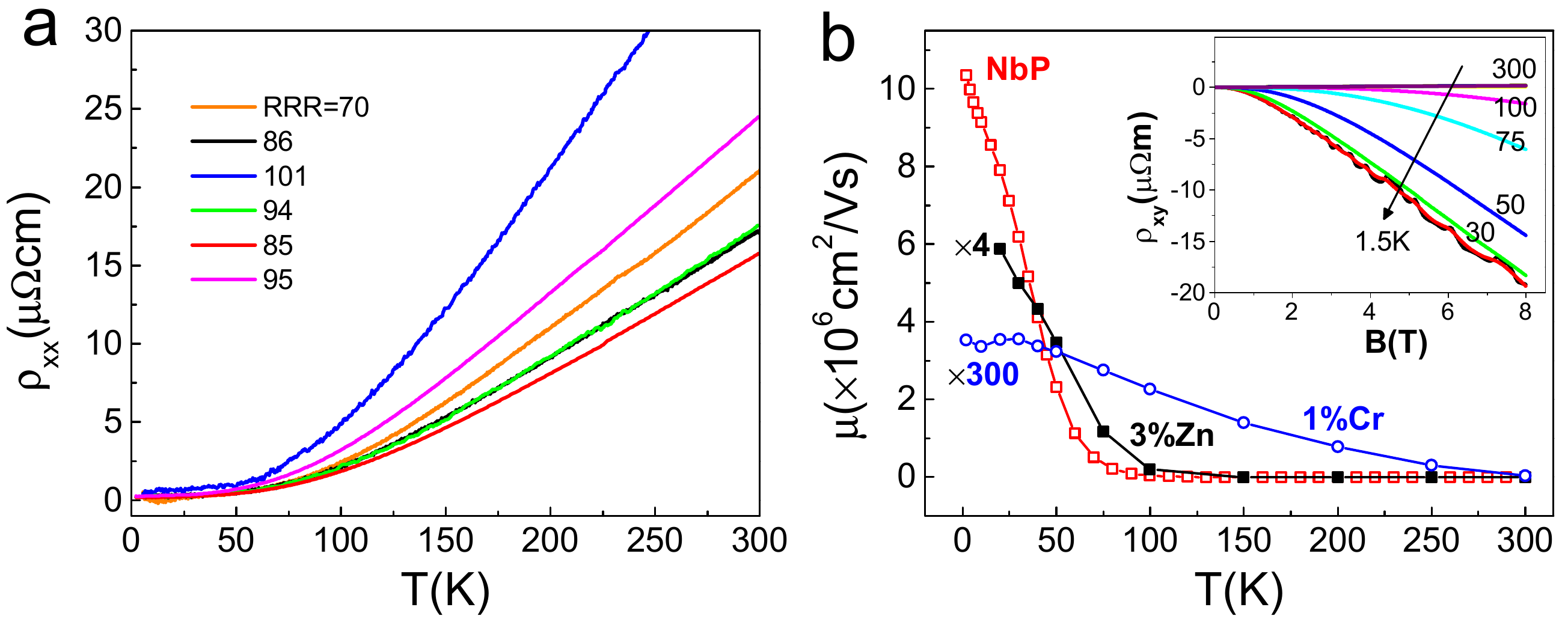}
\end{center}
\caption{\label{Figure1} (a) T-dependent $\rho$ curves at B=0, showing varying sample quality. (b) Ultrahigh carrier mobility of NbP, which is sensitive to magnetic impurities. Inset: T-dependent Hall resistivity from 300 to 1.5 K, showing a pronounced transition from p-type carriers at room temperature to the co-existence of electrons and holes below 200 K.}
\end{figure}

The detailed single crystal growth and structure analysis of NbP are described in the Supplemental Material \footnote{See Supplemental Material at http://link.aps.org/supplemental/ for singlecrystal growth, EDX analysis of elemental compositions, and other supporting transport results.}. Figure \ref{Figure1}a shows the characteristic temperature (T)-dependent resistivity measurements at zero magnetic field (\textit{B}). The results of more than 20 single crystals consistently show decent residual resistance ratio [RRR=$\rho_{xx}(300 K)/\rho_{xx}(1.5 K)$] in the range of 70-100, similar to the previous reports in TaAs and NbAs \cite{TaAs_arXiv_Jia,TaAs_arXiv_NMR,NbAs_jpcm}. The linear $\rho$ above 150 K is typical for metal with dominant electron-phonon (e-ph) scattering. However, the quadratic behavior (T$^n$, $n\sim2.8$) of $\rho$ below 150 K can not be simply explained by e-ph scattering, but indicating limiting scattering mechanism of electron-electron (e-e) interactions. Below 30 K, $\rho$ becomes linear again, which will be correlated to the helicity protection mechanism. Using the single-band theory \cite{HallinMetal_Colin}, we estimate the electron concentration at 1.5 K from Hall signals to be $\sim2\times10^{18}$ cm$^{-3}$ \footnote{The two-band theory fitting of the conductivity tensor $\sigma_{xy}=\rho_{yx}/(\rho_{xx}^2+\rho_{yx}^2)$ suggests the existence of two WSM pockets with opposite carrier polarities. See [25].}. Noticeably, the deduced electron Hall mobility at 1.5 K is very weakly dependent on RRR. With RRR=95, the sample in Fig. \ref{Figure1}b (red squares) has a stunning mobility exceeding $1\times10^{7}$ cm$^{2}$V$^{-1}$s$^{-1}$, while a polished sample with RRR=25 showing $5\times10^{6}$ cm$^{2}$V$^{-1}$s$^{-1}$ \cite{Note1}. The ultrahigh mobility of NbP is comparable to DSM Cd$_{3}$As$_{2}$ \cite{DSMCdAs_NPOng_NatMat15}, despite that RRR is nearly two orders of magnitude higher in the latter. Since RRR is a direct measure of defect concentrations, the observation indicates that the main charge carriers in NbP are effectively protected from defect scattering at zero field. To get insights into the protection mechanism, we have synthesized $\sim$1\% Cr- and $\sim$3\% Zn-doped NbP single crystals \cite{Note1} to study the effects of chemical impurities on the mobility. As shown in Fig. \ref{Figure1}b, with the presence of minimal magnetic impurities, the mobility of Nb$_{0.99}$Cr$_{0.01}$P (blue circles) is degraded by almost three orders of magnitude, while significantly higher concentration of non-magnetic Zn yields comparable mobility to pristine NbP with similar RRR. The chemical doping experiments strongly suggest that the dominant charge carriers in NbP are spin-polarized \cite{WrayTIMagnetic_NatPhy11}, which is consistent with the existence of helical WSM pockets in NbP.

\begin{figure*}[!thb]
\begin{center}
\includegraphics[width=7in]{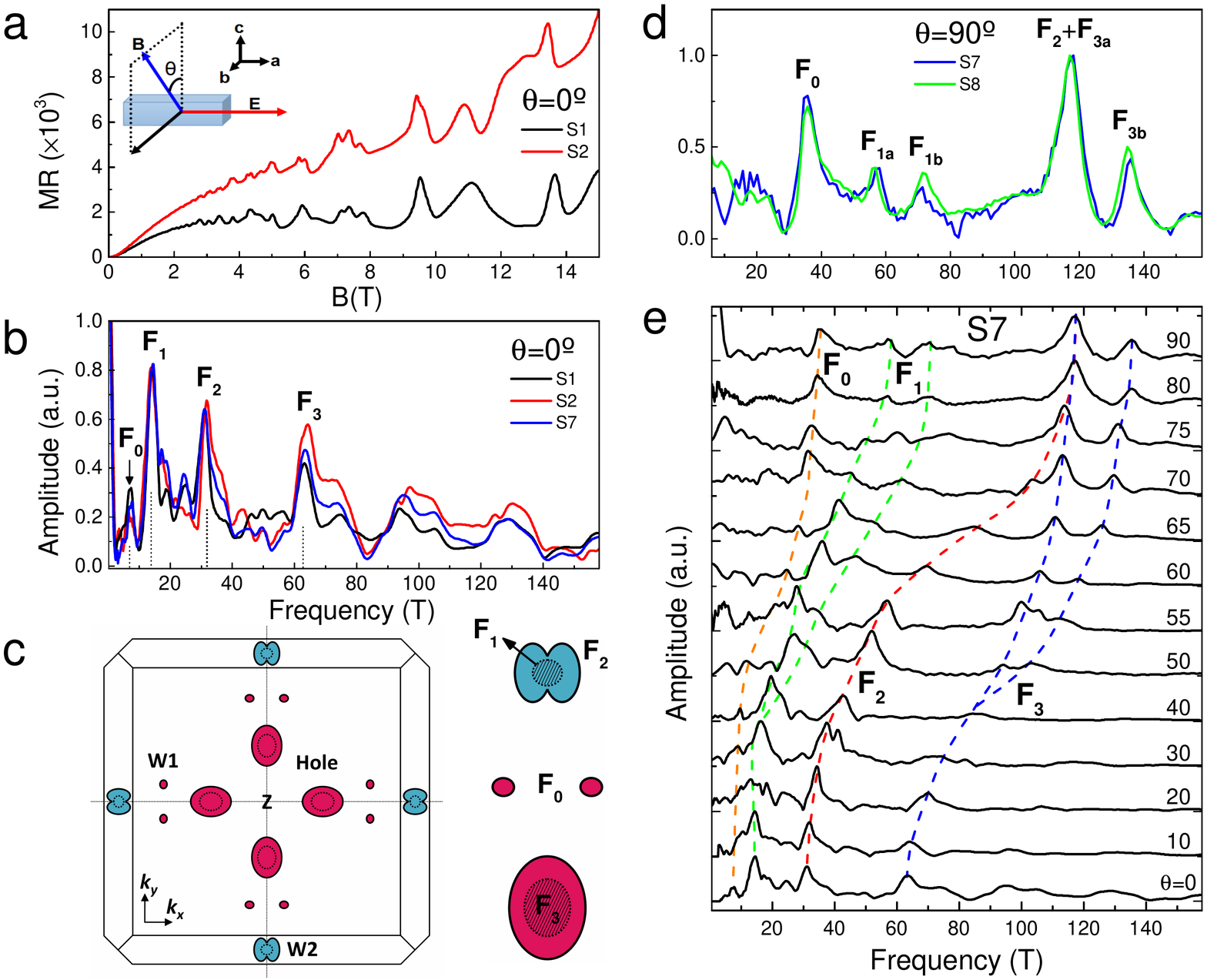}
\end{center}
\caption{\label{Figure2} Angle-dependent SdH oscillations of NbP. (a) Extremely strong SdH oscillations in NbP, with amplitudes independent on RRR. Inset: Schematic of the angle rotation. (b) Four major oscillation peaks of $F_{0}$=6.8 T, $F_{1}$=13.9 T, $F_{2}$=31.8 T and $F_{3}$=64.4 T with B$\|c$. (c) Schematic of FS cross-sections for different pockets with B$\|c$. (d) SdH oscillation peaks with $\theta$=90$^{\circ}$. (e) The evolution of $F_{0}$, $F_{1}$, $F_{2}$ and $F_{3}$ as a function of $\theta$. }
\end{figure*}

We further studied quantum oscillations in NbP to confirm the WSM origin of the spin-polarized carriers. Like the other TaAs family members, NbP has rather complex FSs due to the coexistence of multiple charge carrier pockets. Using density functional theory (DFT) calculations  \cite{Note1}, we found that the FSs of NbP are consisted of p-type W1-WSM, n-type W2-WSM, and eight large travial hole pockets along the $Z-S$ lines. Uniquely, each pair of W2-WSM pockets, enclosing W2 nodes with the opposite chirality, are emerged into a \textit{dumbbell}-shaped continuous FS with an inner FS of trivial electrons, while all W1-WSM pairs are much smaller and well separated in k-space. Experimentally, we have analyzed the FSs of NbP using angle-dependent Shubnikov-de Haas (SdH) oscillations. Different from the literatures, we rotated $B$ in perpendicular to the electric field ($E$), and thus define the rotation angle ($\theta$) by the orientation of $B$ and the $c$-axis (Fig. \ref{Figure2}a). Such configuration not only allows us to get robust SdH oscillations for all $\theta$ setpoints, but also excludes possible signals from chiral anomaly.

With $\theta$=0$^{\circ}$ ($B\|c$), pristine NbP at 2 K is characterized by unusually large magnetoresistance, MR=$[\rho_{xx}(B)-\rho_{xx}(0)]/\rho_{xx}(0)$, and extremely strong SdH oscillations (Fig. \ref{Figure2}a). The SdH peaks become visible once B exceeding 0.5 T, and all measured crystals show non-saturating quasi-linear MR, which can reach striking 10,000 at 15 T. Similar to the electron mobility, the amplitudes of the SdH oscillations are also weakly correlated to RRR, while the linear MR is rather sample dependent (Fig. \ref{Figure2}a and \cite{Note1}). This also implies that the SdH oscillations are controlled by the spin-polarized carriers with ultrahigh mobility, while the linear MR may require compensation mechanism \cite{WTe2NPOng_Nature14} from the large, low-mobility hole pockets. Indeed, the chemical doping of Zn greatly suppresses RRR, but the SdH oscillations remain robust in Nb$_{0.97}$Zn$_{0.03}$P. In contrast, quantum oscillations are completely absent in Nb$_{0.99}$Cr$_{0.01}$P.

By performing fast Fourier transformation (FFT), we get four major oscillation frequencies of $F_{0}$=6.8 T, $F_{1}$=13.9 T, $F_{2}$=31.8 T and $F_{3}$=64.4 T, respectively (Fig. \ref{Figure2}b). Since $B$ is applied in the direction with the four-fold rotation symmetry, the same type of carrier pockets are quantized equivalently. In this case, SdH oscillations detect the FS cross-sections of different carrier pockets in the $k_{x}$-$k_{y}$ plane (Fig. \ref{Figure2}c). By comparing the experimental oscillation frequencies to the DFT calculations, we can correlate the FFT peaks to the existing pockets, as summarized in Table \ref{Oscillations}. Note that the $F_{3}$ trivial hole pocket is surrounded by a much larger hole FS of 133 T, thus forming the complex inner and outer FSs similar to the case of $F_{1}$ and $F_{2}$ \cite{BerryPhaseRashba_Science}. The assignment of FFT peaks have also been cross-checked by complementary magnetic oscillations of Nernst and Seebeck coefficients as well as the de Haas-van Alphen (dHvA) effect \cite{Note1}. As shown in Table \ref{Oscillations}, all three techniques quantitatively agree with the SdH results.

\begin{table}[h]
\caption{Quantum oscillation frequencies (T) determined by experiments and DFT with $B\|c$.}
\resizebox{3.5in}{!}{
\begin{tabular}{|c|c|c|c|c|c|c|}
\hline
\multirow{2}{*}{Pockets} & \multirow{2}{*}{\textbf{W1-WSM($F_{0}$)}} & \multicolumn{2}{c|}{\textbf{W2-WSM}} & \multicolumn{2}{c|}{\textbf{Hole}} \\ \cline{3-6}
                         &               & Inner\textbf{($F_{1}$)} & Outer-WSM\textbf{($F_{2}$)}             & Inner\textbf{($F_{3}$)}       & Outer           \\ \hline
DFT                      & 4.1 (Hole)    & 12.5          & 35.6            & 66.2         & 133.8    \\ \hline
SdH                      & 6.8           & 13.9          & 31.8            & 64.4         & 130.1    \\ \hline
Nernst                   & 6.9           & 14.0          & 31.7            & 63.8         & 127.3    \\ \hline
Seebeck                  & 7.1           & 12.2          & 31.6            & 61.7         & 123.8    \\ \hline
dHvA (5T)                & 7.3           & 12.3          & 31.0            & 62.0         & N.A.
\\ \hline
\end{tabular}
}
\label{Oscillations}
\end{table}

Using angle- and T-dependent SdH oscillations \cite{MagOsi_Shoenberg}, we determined indispensable information on the anisotropy of the individual FS and the effective mass ($m^{*}$) of the corresponding carriers, respectively. As shown in Fig. \ref{Figure2}d and \ref{Figure2}e, we have observed dramatic $\theta$-dependent changes in $F_{2}$, which monotonically shifts up to 118 T when $\theta$ is increased from 0$^{\circ}$ to 90$^{\circ}$ (the red dashed line in Fig. \ref{Figure2}e), agreeing with the highly anisotropic nature of the W2-Weyl cones. In contrast, $F_{0}$ gradually reaches a maximum frequency of 35 T at 90$^{\circ}$. Such isotropic FS is also expected for the W1-WSM pockets, which are dates-like ellipsoids \cite{TaAsDingH_ARPESNode,TaAs_arXiv_NMR}. Noticeably, the trivial pockets of $F_{1}$ and $F_{3}$ are both splitting into two oscillation peaks when $\theta>30^{\circ}$, which are typical for ellipsoid-shaped FSs of parabolic energy bands. The lower frequency part of $F_{3}$ is coincident with the W2-WSM peak to form the dominant peak of 118 T at $\theta$=90$^{\circ}$.

\begin{figure}[!thb]
\begin{center}
\includegraphics[width=3.5in]{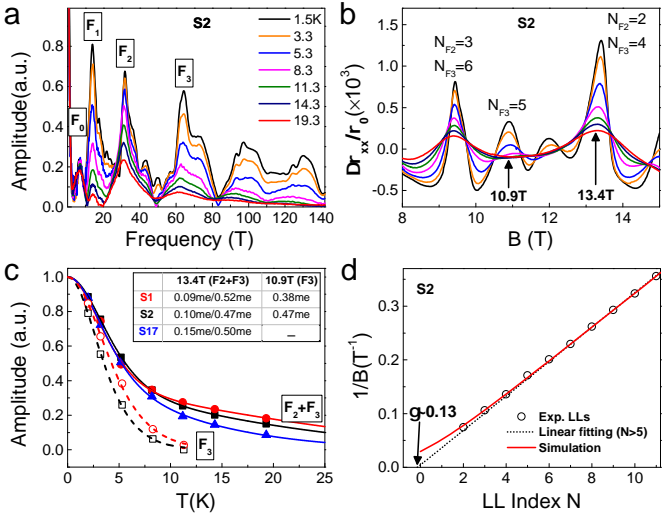}
\end{center}
\caption{\label{Figure3} T-dependent SdH oscillations and Berry's Phase. (a) FFT of T-dependent SdH oscillations of S2. (b) T-dependent SdH oscillations in the field range of 8-15 T. (c) The LK model fitting of the 10.9 T and 13.4 T peaks. The latter is the superposition of $F_{2}$ and $F_{3}$. (d) Deviation of Berry's phase from $\pi$ due to  non-ideal relativistic fermions in the W2-WSM pockets.}
\end{figure}

For WSM pockets, the linear energy dispersion results in significantly smaller $m^{*}$ for Weyl fermions, compared to trivial pockets. Taking the quantization condition of $\hbar \omega_{c}/k_{B}T \geq 1$, in which $k_{B}$ is the Boltzmann’s constant and $\omega_{c}$=$eB/m^{*}$ is the cyclotron frequency, we can expect the SdH oscillations of WSM pockets to be persistent at much higher T than the trivial ones. Indeed, $F_{0}$ and $F_{2}$ remain robust at 20 K, while $F_{1}$ and $F_{3}$ are not discernable above 10 K (Fig. \ref{Figure3}a). We have determined the effective mass $m^{*}$ for $F_{2}$ and $F_{3}$, using the Lifshitz-Kosevich (LK) formula for 3D systems:
\begin{equation}\label{eq1}
  A(B,T) \propto \exp(-\frac{2\pi^{2}k_{B}T_{D}}{\hbar \omega_{c}})\frac{2\pi^{2}k_{B}T/\hbar \omega_{c}}{\sinh (2\pi^{2}k_{B}T/\hbar \omega_{c})},
\end{equation}
where A(B,T) is the SdH amplitude, and $T_{D}$ is the Dingle temperature. As shown in Fig. \ref{Figure3}b, there are four prominent SdH peaks in the field range of 8-15 T. The 9.4 T and 13.4 T peaks are the superimposition of the N=3 and N=2 Landau levels (LLs) of $F_{2}$ and the N=6 and N=4 of $F_{3}$, respectively. The fitting of 9.4 T and 13.4 T peaks requires two distinct effective mass of $m_{F_{2}}^{*}$=0.1$m_{e}$ and $m_{F_{3}}^{*}$=0.47$m_{e}$, respectively (Fig. \ref{Figure3}c). For the 10.9 T peak, which is solely contributed by $F_{3}$ (N=5), single LK fitting yields 0.45m$_{e}$, agreeing well with the double LK fitting. For $F_{0}$ and $F_{1}$, it is difficult to extract $m^{*}$ directly from MR oscillation peaks due to the superimposition of much higher frequencies of $F_{2}$ and $F_{3}$. Instead, we analyzed the T-dependent FFT amplitudes and got $m_{F_{0}}^{*}$=0.06$m_{e}$ and $m_{F_{1}}^{*}$=0.29$m_{e}$ \cite{Note1}. Like $F_{3}$, $F_{1}$ also becomes vanishingly small when T is above 10 K, supporting its origin in the inner trivial electron FS of the W2-WSM pockets. Distinctively, the SdH oscillations of NbP at 20 K are characterized by strong second harmonic peaks of $F_{0}$ and $F_{2}$, a manifestation of low $m^{*}$ and small $T_{D}$ of Weyl fermions \cite{Note1}.

Nontrivial Berry's phase of $\pi$ is the quantum signature of the liner energy bands \cite{BerryPhase_PRL99} in DSMs and WSMs. We have determined the $\Phi_{B}$ of the W2-WSM state, using the Landau fan diagram extracted from the SdH peaks from six different samples. Surprisingly, a simple linear fitting with the constraint frequency of $F_{2}$ gives an odd Onsager phase of $\gamma\sim 0.3$ \cite{Note1}, which is considerably exceeding the geometrical correction factor for a 3D FS, following the Lifshitz-Onsager relation of $\gamma=1/2-\Phi_{B}/2\pi+\delta$ \cite{BerryPhaseRashba_Science,3DBerryPhase_PRL04}. A detailed examination of the experimental data reveals that for high LL index $N>5$, linear fitting of the the LL fan diagrams produces $\gamma\sim 0.13$, which gradually shifts to the odd number of $0.3$ as $N$ approaching the ultra-quantum limit. Such deviation is a manifestation of the non-ideal relativistic fermions \cite{Ando_StrangeBerryPhase_PRB} in the W2-WSM pockets, which have non-negligible $m^{*}_{F_{2}}=0.1 m_{e}$ and relative low Fermi velocity of $v_{F}\sim 1.8\times10^{5} m/s$. Using the method proposed by Taskin \textit{et al} \cite{Ando_StrangeBerryPhase_PRB}, we are able to simulate the systematic changes in $\gamma(N)$, as shown by the red solid line in Fig. \ref{Figure3}d.

\begin{figure}[!thb]
\begin{center}
\includegraphics[width=3.5in]{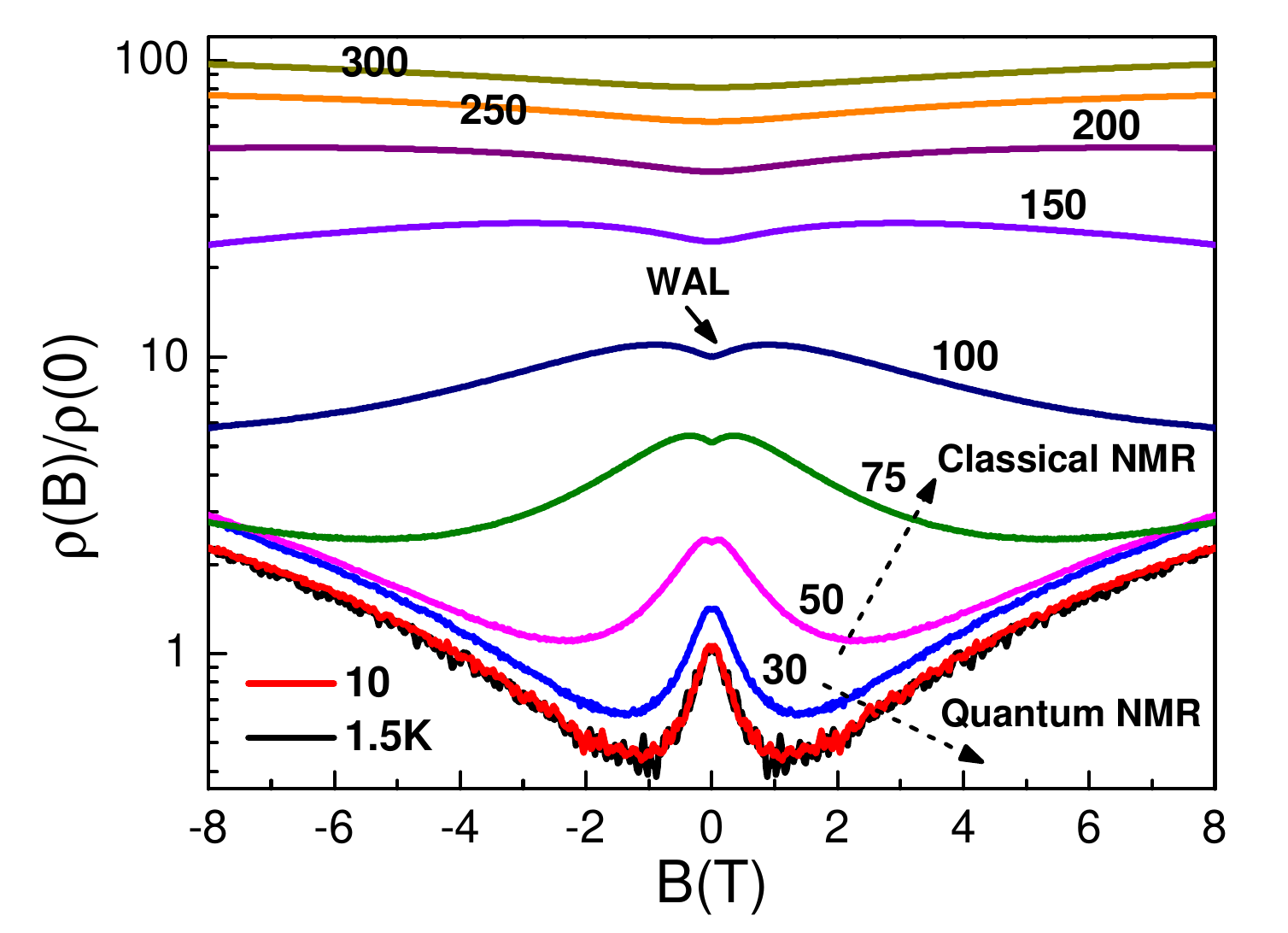}
\end{center}
\caption{\label{Figure4} Robust chiral anomaly induced NMR in NbP as a function of T, showing T-independent NMR below 10K and strong T dependence above 30 K.}
\end{figure}

With the presence of chiral Weyl node pairs in NbP, NMR induced by the Adler-Bell-Jackiw anomaly \cite{ChiralAnomaly_PLB} would be expected when $B\|E$. Surprisingly, we have observed very robust NMR in NbP far above the quantization temperature regime. As shown in Fig. \ref{Figure4}, the MR curves form a sharp negative dip below 2 T with $T<$50 K. The overall MR change at 1.5 K is about -80\%, in contrast to -30\% reported in TaAs \cite{TaAs_arXiv_NMR}. The NMR becomes weaker and broader when $T$ increases, but persists up to 150 K. Considering that each W2-WSM pockets are continuous FSs enclosing one W2 pair, an extra chiral current channel between different W2 nodes is not allowed. In contrast, each pair of the W1-WSM pockets are well separated in k-space. Using $F_{0}$ and $m^{*}=0.06\,m_{e}$, we estimated the $E_{F}$ is $\sim 15$ meV below the W1 nodes. By taking the thermal activation energy of 13 meV (150 K) into account, the results qualitatively agree with the DFT calculations which suggest a dome structure of 25 meV below the W1 nodes along the internode direction. It is distinctive that the NMR effects are identical below 20 K. In this quantum regime, the longitudinal chiral conductivity is expressed by
\begin{equation}\label{eq2}
 \triangle \sigma_{c}\propto\frac{B^{2}v_{F}^3}{E_{F}^2{}}\tau(E_{F}),
\end{equation}
where $\tau(E_{F})$ is the elastic intervalley scattering time by impurities \cite{Spivak_NMRModeling}. In this regime, Weyl fermions are strictly protected by helicity from the scattering of non-magnetic impurities. It leads to large constant $\tau$ and thus identical NMR behavior. At elevated temperatures above 20 K, however, the intravalley inelastic scattering between electrons cannot be ignored, and the chiral anomaly induced conductivity enhancement becomes $\triangle \sigma_{c}\propto\frac{B^{2}v_{F}^3}{T^2{}}\tau_{T}$, in which the T-dependent $\tau_{T}$ is modified by electron-electron inelastic scattering $\tau_{e}$ \cite{Spivak_NMRModeling}. Such T-dependent NMR behavior is consistent with the zero-field resistivity measurements in Fig. \ref{Figure1}a.

Our discovery not only proves the existence of exotic WSM states in NbP, but also provides unambiguous evidence in correlating the ultrahigh mobility to the spin conservation of helical Weyl fermions. Unlike the pseudo-spin in graphene, which is vulnerable to lattice defects and atomic-scale disorders, the helical spin-textures in NbP are topologically protected by the non-centrosymmetric symmetry, thus, are remarkably robust against non-magnetic disorders. For pristine NbP, the doping is nearly intrinsic despite that there are small amounts of excessive P in crystals (P:Nb=50.5:49.5$\pm$3\% \cite{Note1}). Nevertheless, a strategy to continuously tune the doping level in NbP would be highly desirable, and may open enormous opportunities for exploring various topological quantum phenomena and spin device concepts.

\begin{acknowledgments}
We thank Yayu Wang, Kai Wang and Fuchun Zhang for insightful discussions. This work was supported by the National Basic Research Program of China (Grant Nos. 2014CB921203 and 2012CB927404), the National Science Foundation of China (Grant Nos. 11190023, U1332209 and 11574264), and the Fundamental Research Funds for the Central Universities of China. Y.Zheng acknowledges the start funding support from the 1000 Youth Talent Program.
\end{acknowledgments}

\end{document}